\documentclass[aps,showpacs,twocolumn,amsmath,amssymb]{revtex4}
\usepackage{graphicx}
\usepackage{dcolumn}
\usepackage{bm}
\begin{document}

\title{Use strong coupling strength to coherently preserve quantum entanglement}
\author{Guihua Tian}\email{tgh-2000@263.net, tgh20080827@gmail.com }

  \affiliation{School of Science, Beijing
University of Posts And Telecommunications. Beijing 100876 China.}
\author{Shuquan Zhong}\email{shuqzhong@gmail.com}
 \affiliation{State Key Laboratory of Information Photonics and Optical
Communications, \\ Beijing University of Posts And
Telecommunications.
 Beijing 100876 China.}

\begin{abstract}
The dynamics of two qubits ultra-strongly coupled with a quantum
oscillator is investigated by the adiabatic approximation method.
The evolution formula of the  initial four Bell states are studied
under the control mechanism of the coherent state of the quantum
oscillator. The influential parameters for the preservation of the
entanglement are the four parameters: the average number of the
coherent state, the ultra-strong coupling strength, the ratio of two
frequencies of qubit and oscillator, and the inter-interaction
coupling of the two qubits. The novel results show that the
appropriate choice of these parameters can enable this mechanism to
be utilized to preserve the entanglement of the two qubits, which is
initially in the state $|I_0\rangle$ of the four Bell states. We
give two different schemes to choose the respective parameters to
maintain the entangled state $|I_0\rangle$ almost unchanged. The
results will be helpful for the quantum information process.
\end{abstract}

 \maketitle

\section{Introduction}
Entanglement is indispensable in the quantum information process.
Quantum entanglement states have been applied in quantum key
distribution and teleportation, entanglement purification,
factorization of integers, random
searches\cite{chuang}-\cite{ekert}. Generation and preservation of
entanglement of qubits are crucial for all the quantum information
process.  It is still challenge to externally control entanglement.
Recently, using quantum bus to coherently and controllably
manipulate quantum entanglement is provided in Ref.\cite{m}, where
the famious Jaynes-Cummings (JC) model is applied as the control
mechanism.

Theoretically, the entanglement generation and maintaining of two
qubits can be achieved by use of Jaynes-Cummings (JC) model
Hamiltonian\cite{yu}.  The entanglement reciprocation is also
studied  between the field variables  and a pair of qubitsin JC
cavity\cite{krau}-\cite{zhou}. Later, study shows that
coherent-state control of a pair of non-local atom-atom entanglement
between two spatially separated sites is possible\cite{m}. There are
 some time-dependent entanglement death and rebirth effects in these investigations.

 JC model is a main mechanism to be used to study how to control the
 manipulation of quantum qubits, whose validity rely on the
 assumption of the weak coupling of quantum oscillator, qubits and their near
resonance condition. It is obtained from the Rabi model  by
discarding the count-rotating-wave terms\cite{mand}.  The strong and
ultra-strong coupling region of qubit and
 oscillator  provide many new and anti-intuitive
results for the Rabi model\cite{iris}-\cite{asha}, which   is
treated in Ref.\cite{sorn} with zero detuning. Further, ultra-strong
coupling and large detuning is investigated, and novel results
appear, like frequency modification, collapse and revival of Rabi
oscillation for one qubit with the initial state of the single mode
field being thermal or coherent\cite{iris}-\cite{asha}. In
Ref.\cite{agar}, Rabi model is extended to two qubits case, where
the authors studied the death and revival phenomena for the two
qubits' entanglement for the initial coherent state of the
oscillator. However, they only study very special case of the
coupling parameter $\beta^2\ll \Omega_{1N}$  (see Ref.\cite{agar} or
the following for details), which will greatly simplify the
eigenvectors and subsequent calculation. There are no investigation
concerning the range where the coupling strength does not satisfy
$\beta^2\ll \Omega_{1N}$.

Enlightened by these works, we investigate use the Rabi mechanism
control. In our study, we will not restrict our study by the
condition  $\beta^2\ll \Omega_{1N}$, and this in turn will make
calculation complicated. Never the less, it also give one chance to
obtain some unexpected phenomena, that is the new way to  use
coherent quantum mode to control one of the full entangled Bell
state and novel results are obtained based on the complex
calculation, which will helpful for the quantum information process.

In addition, the condition $\beta^2\ll \Omega_{1N}$ in
Ref.\cite{agar} is not easy to be satisfied due to the fact
$\Omega_{1N}$ depends on the parameters $N,\ \beta,\
\frac{\omega_0}{\omega}$ nonlinearly. So the investigation without
the condition is crucial for their further application in quantum
information process. Furthermore, the above mentioned complexity
unexpectedly promotes our ability to preserve the entanglement of
the two qubits in the state $|I_0\rangle$, one of the four Bell
states. This unexpected result can be exploited in quantum
information process involved the entanglement of two qubits.

 The paper organized as
follows: we first introduce the two-qubit system with inter-qubit
coupling briefly and give a simple model for it. Then we give a view
of the method to to be used to study the Rabi model in ultra-strong
coupling regime with lager detuning, the adiabatic approximation
method (AA). The spectrum of the qubits coupled with quantum mode
filed is given in subsequent section, Then the evolution of the
system are investigated and followed by the section to study the
preservation of the entanglement.

\section{The two qubits with inter-qubit coupling}
Because we will treat the two qubits system as three-level-energy
system, which  will be used both for circuit QED and cavity QED.
Here we give a brief introduction about it.   The Hamiltonian for
the system of a two-qubit  with inter-qubit
coupling\cite{jing}-\cite{fice}
\begin{eqnarray}\label{H2 for qutrit}
     H_{q}= \frac12\hbar\omega_0(\sigma_z^{(1)}+\sigma_z^{(2)})
     +\kappa\hbar\omega_0(\sigma_-^{(1)}\sigma_+^{(2)}+\sigma_-^{(2)}\sigma_+^{(1)}),
     \end{eqnarray}
where $ \sigma_z^{(i)}, \ i=1,2$ are the ith qubit's Pauli matrices
and $\kappa$ is the coupling strength between the qubits.
$\sigma_{\pm}^{(i)}$ are\begin{eqnarray} \sigma_{+}^{(i)} =\left(
          \begin{array}{ccc}
           0 & 1 \\
                  0& 0 &\\
          \end{array}
        \right),\ \
\sigma_{-}^{(i)}=\left(
          \begin{array}{ccc}
           0 & 0 \\
                  1& 0 &\\
          \end{array}
        \right)
\end{eqnarray}
for $ith$-qubit.  The Hamiltonian could be a diagonal matrix under
the collective states \[|3\rangle=|\uparrow \uparrow \rangle,\]\[
|s\rangle=\frac1{\sqrt2}(|\uparrow \downarrow\rangle+|\downarrow
\uparrow\rangle),\]\[ |a\rangle=\frac1{\sqrt2}(|\uparrow
\downarrow\rangle-|\downarrow \uparrow\rangle),\]\[
|1\rangle=|\downarrow \downarrow \rangle\] as basis
\cite{jing}-\cite{dick}:
\begin{eqnarray}
H'_q=\hbar\omega_0\left(
      \begin{array}{cccc}
       1 & 0 & 0 &0 \\
                 0 & \kappa & 0 &0 \\
                 0 & 0 & -\kappa&0 \\
                 0 & 0 & 0& -1 \\
      \end{array}
    \right).
\end{eqnarray}
There are two transition channels, the symmetric one
$|3\rangle\rightarrow|s\rangle\rightarrow |1\rangle $
 and the asymmetric one $|3\rangle\rightarrow|a\rangle\rightarrow
|1\rangle$. The two channels  are not correlated
\cite{jing}-\cite{dick}, so we will divide the $H_q $ as

\begin{eqnarray}
H_q=\hbar\omega_0\left(
      \begin{array}{cccc}
       1 & 0 & 0  \\
                 0 & \kappa & 0  \\
                 0 & 0 &  -1 \\
      \end{array}
    \right)
\end{eqnarray}
for symmetric transition channel or
\begin{eqnarray}
H_q=\hbar\omega_0\left(
      \begin{array}{cccc}
       1 & 0 & 0  \\
                 0 & -\kappa & 0  \\
                 0 & 0 &  -1 \\
      \end{array}
    \right)
\end{eqnarray}
for asymmetric channel transition. We will unite the two cases as
\begin{eqnarray}
H_q=\hbar\omega_0\left(
      \begin{array}{cccc}
       1 & 0 & 0  \\
                 0 & a & 0  \\
                 0 & 0 &  -1 \\
      \end{array}
    \right)
\end{eqnarray}
with $a$ positive and negative for symmetric and asymmetric channels
respectively. We also denote $|2\rangle=|s\rangle$ for symmetric
channel and $|2\rangle=|a\rangle$ for asymmetric channel
respectively.

The interacting two identical qubits will correspond to two
three-level-energy systems with the same top and bottom eigenstates
and different middle states. The top and bottom states' energies are
$\hbar\omega_0, \ -\hbar\omega_0$. The two middle states will have
the same energy if the two qubits do not couple with each other. In
this case, the two three-level-energy systems are the same
concerning their energy distributions. Generally, they will be
regarded as just one. However, the two middle states will have
different energies, one positive and the other negative. The
corresponding two three-level-energy systems are different even
concerning with their energy distribution. Nevertheless, relating
the transition, the two three-level-energy systems are not
correlated with each other, so we could consider one .

\section{The Rabi Hamiltonian }

As the Hamiltonian of the two qubits can be represented a $3\times
3$ matrix, the Rabi model is extended as describing the dynamics of
the two qubits interacting with a single quantum mode field
by\begin{eqnarray}\label{H for qubits}
     H=\hbar\omega_{0}S_z+ \hbar\omega a^{\dagger}a
+ \hbar\omega\beta(a+a^{\dagger})S_x , \end{eqnarray}  which will be
the Rabi model as $S_x,\ S_z$ reduce to the usual Pauli matrices.
The qubits is described by $H_q=\hbar\omega_0$  before stated.In the
matrix form, $S_x,\ S_z$ are
\begin{eqnarray}
S_x=\left(
      \begin{array}{ccc}
       0 & 1 & 0 \\
                 1 & 0 &1 \\
                 0 & 1 & 0 \\
      \end{array}
    \right),
S_z=\left(
      \begin{array}{ccc}
         1 & 0 & 0 \\
                 0 & a &0 \\
                 0 & 0 & -1\\
      \end{array}
    \right).
\end{eqnarray}
The operator $S_x$ is connected with the operators $\sigma^+,\
\sigma^-$
\[S_x=\sigma^++\sigma^-,\]
where\begin{eqnarray} \sigma^+=\left(
      \begin{array}{ccc}
       0 &0 & 0 \\
                 1 & 0 &0 \\
                 0 & 1 & 0 \\
      \end{array}
    \right),\
\sigma^-=\left(
      \begin{array}{ccc}
         0 & 1 & 0 \\
                 0 & 0&1 \\
                 0 & 0 & 0\\
      \end{array}
    \right).
\end{eqnarray}
The eigenstates $|1\rangle, \ |2\rangle, |3\rangle $ change under
the operators $\sigma^+,\ \sigma^-$ as following
\[\sigma^+|1\rangle=|2\rangle,\  \sigma^+|2\rangle=|3\rangle ,\sigma^+|3\rangle=0,\]\[\sigma^-|1\rangle=0, \
\sigma^-|2\rangle=|1\rangle, \ \sigma^-|3\rangle=|2\rangle.\]
 As stated before, the
Hamiltonian of the system is not analytically integrable. The RWA
proximation supposes the resonate condition $\omega_0\approx \omega$
and the weak coupling $\beta\ll 1$ and becomes completely solvable
by discarding the non-energy conserving terms $a\sigma^-,\
a^{\dagger}\sigma^+$.

\section{Adiabatic Approximation in Ultra-strong Coupling Range}

Whenever the coupling is strong or the detuning is large, the
counter-RWA terms $a\sigma^-,\ a^{\dagger}\sigma^+ $ can not be
omitted. This belongs to the regime of adiabatic approximation . In
adiabatic approximation, $\omega_0 $ is small relative to the other
terms in the Hamiltonian, and one could first omit it to study the
rest as the non-RWA Hamiltonian , then take it as perturbation in
later. Physically, this focuses on the quantum oscillator influenced
by the term $\hbar\beta(a+a^{\dagger})S_x$. The Hamiltonian reads
\begin{eqnarray}
 H^{0}= \hbar\omega a^{\dagger}a
+ \hbar\beta(a+a^{\dagger})S_x.
\end{eqnarray}
If studied classically, the oscillator undergoes some forced motion
by the qubits. The quantum oscillator system interacting with one
qubit and two non-interacting qubits have been solved in the
adiabatic approximation (see references\cite{iris}-\cite{agar}). We
now employ the similar method to solve the non-equal-level qubits
system. The eigen-vecrtors  $ |1,1 \rangle,\ |1, 0\rangle,\ |1,-1
\rangle $ of the operator $S_x $ are\begin{eqnarray}
S_x|1,m\rangle=\sqrt2m |1,m\rangle,\ m=0,\pm1,
\end{eqnarray} which are written as
\begin{eqnarray} \left(
  \begin{array}{c}
    |1,1 \rangle \\ |1, 0\rangle \\
  |1,-1 \rangle\\
  \end{array}
\right)=\left(
          \begin{array}{ccc}
           1/2 & 1/\sqrt{2} & 1/2 \\
                  1/\sqrt{2} & 0 &- 1/\sqrt{2} \\
                 1/2 & -1/\sqrt{2} & 1/2 \\
          \end{array}
        \right)
\left(
          \begin{array}{c}
            |3 \rangle \\ |2\rangle \\
  |1 \rangle\\
          \end{array}
        \right)\label{10,11,1-1vecctor}.
\end{eqnarray}
With the help of these vectors, the eigen-vectors $|\Psi\rangle$ for
the eigenstates of operator $H^0$ will be written as \cite{iris}
,\cite{agar}
\[|\Psi_{n,m}\rangle=|1,m\rangle|N_m\rangle=|1,m\rangle D(-\sqrt2
m\beta)|N\rangle\] with the corresponding eigen-values
$E^0_{n,m}=\hbar\omega(N-2\beta^2m^2)$. The displaced operator
$D(\alpha)$ of the quantum oscillator is defined as
$D(\alpha)=\exp{(\alpha a^{\dagger}-\alpha^*a)}$ for the arbitrary
complex number $\alpha$. The interaction with the qubits makes
potential well of the quantum oscillator displaced
 according to the states of the qubits. From the physical view,
 the interaction term $\hbar\beta(a+a^{\dagger})S_x$ has influence
 to displace the equilibrium position of the oscillator to different
 points by the different states of the qubits $|1,m\rangle,\ m=0,\pm1$
, which result in three displaced number vectors $|N_m\rangle,\
m=0,\pm 1$. From the mathematical view, the eigenstates
$|1,m\rangle|N_m\rangle, m=0,\pm1,\ N=0,1,2,\cdots $ constitute as a
complete basis for the composite system and are useful for the later
calculation, that is, any vector for the composite system of the
qubits and oscillator could be decomposed in the basis
$|1,m\rangle|N_m\rangle, m=0,\pm,\ N=0,1,2,\cdots $. This basis is
not orthogonal due to the fact
\begin{eqnarray}
\langle N_{l}|M_r\rangle\ne 0 \ for \ r\ne l.
\end{eqnarray}
In order to obtain the spectrum for the Hamiltonian $H$, we need to
make calculation of the terms $\langle 1,l|S_z|1,r\rangle\langle
N_{l}|M_r\rangle, \ l,\ r=0,\pm 1 $. Due to the fact $\omega_0\ll
\omega$, the transition of the qubits generally contributes little
in exciting the quantum oscillator, so the corresponding terms
$\langle 1,l|S_z|1,r\rangle\langle N_{l}|M_r\rangle, \ N\ne M$ could
be omitted. This approximation is called adiabatic approximation
(AA).

\section{The spectrum of the Hamiltonian by AA method} Under
this adiabatic approximation, the Hamiltonian $H$ becomes
block-diagonal with the nth diagonal block $\tilde{H}_N$ as a
$3\times 3$ matrices
 defined under the basis
$|1,m\rangle |N_m\rangle,\ m=1,0,-1$ as
\begin{eqnarray}
\tilde{H}_N=\left(
      \begin{array}{ccc}
        \tilde{N} & \Omega_{1N} &\Omega_{2N} \\
        \Omega_{1N} & N &\Omega_{1N} \\
        \Omega_{2N} & \Omega_{1N} &\tilde{N} \\
      \end{array}
    \right),
\end{eqnarray}
where
\begin{eqnarray}
  \tilde{N} &=& N-2\beta^2+\frac{a}2 \frac{\omega_0}{\omega}, \\
  \Omega_{1N} &=& \frac{\omega_0}{\omega}\langle 1,1|S_z|1,0\rangle\langle N_{1}|N_0\rangle\nonumber \\
  &=&  \frac{1}{\sqrt2}\frac{\omega_0}{\omega}\exp{(-\beta^2)}L_N(2\beta^2) ,\\
  \Omega_{2N} &=& \frac{\omega_0}{\omega}\langle 1,-1|S_z|1,1\rangle\langle N_{-1}|N_1\rangle\nonumber \\
    &=&  -\frac{a}2
    \frac{\omega_0}{\omega}\exp{(-4\beta^2)}L_N(8\beta^2).
\end{eqnarray}
The parameter $a$ enters $\tilde{H}_N$  contributing two diagonal
terms in $\tilde{N}$ and two off-diagonal ones in $\Omega_{2N}$,
which will give rise to the transition between $|1,-1\rangle
|N_{-1}\rangle$ and $|1,1\rangle |N_{1}\rangle$. This is a new
transition due to the non-equal-level parameter $a\ne 0$ and is
absent in the equal-level case. The solutions to the eigen-value
problem of the operator $\tilde{H}_N$ are
\begin{eqnarray}
  \tilde{E}_{N,0}^0&=& \hbar\omega\left(N-2\beta^2+\frac{a}2 \frac{\omega_0}{\omega}-\Omega_{2N}\right)\nonumber \\
  &=& \hbar\omega\left(N+\tilde{T}_0-2\Omega_{2N}\right)\label{e0}, \\
  \tilde{E}^0_{N,\pm} &=& \hbar\omega\left(N+\frac{\tilde{T}_0\pm \sqrt{\tilde{T}_0^2+8\Omega^2_{1N}}}2\right),\label{epm} \\
  \tilde{T}_0 &=&  -2\beta^2+\frac{a}2 \frac{\omega_0}{\omega}+\Omega_{2N} \label{t0}
  \nonumber\\ &=&-2\beta^2+\frac{a}2 \frac{\omega_0}{\omega}(1-\exp{(-4\beta^2)}L_N(8\beta^2)),
\end{eqnarray}
and
\begin{eqnarray}
  |\tilde{E}_{N,0}^0\rangle &=& \frac1{\sqrt2}\left(
                      \begin{array}{c}
                        1 \\
                        0 \\
                        -1 \\
                      \end{array}
                    \right), |E^0_{N,\pm}\rangle =\frac1{\tilde{L}_{N,\pm}}\left(
                      \begin{array}{c}
                        1 \\
                        \tilde{Y}_{N,\pm} \\
                        1 \\
                      \end{array}
                    \right),  \nonumber \\
                    \label{vpm0}\\
  \tilde{Y}_{N,\pm} &=& \left(\frac{-\tilde{T}_0\pm \sqrt{\tilde{T}_0^2+8\Omega^2_{1N}}}{2\Omega_{1N}}\right), \\
  \tilde{L}^2_{N,\pm} &=&  \tilde{Y}^2_{N,\pm} +2.
\end{eqnarray}
\begin{figure}[ht]
\centering
\includegraphics[width=0.38\textwidth]{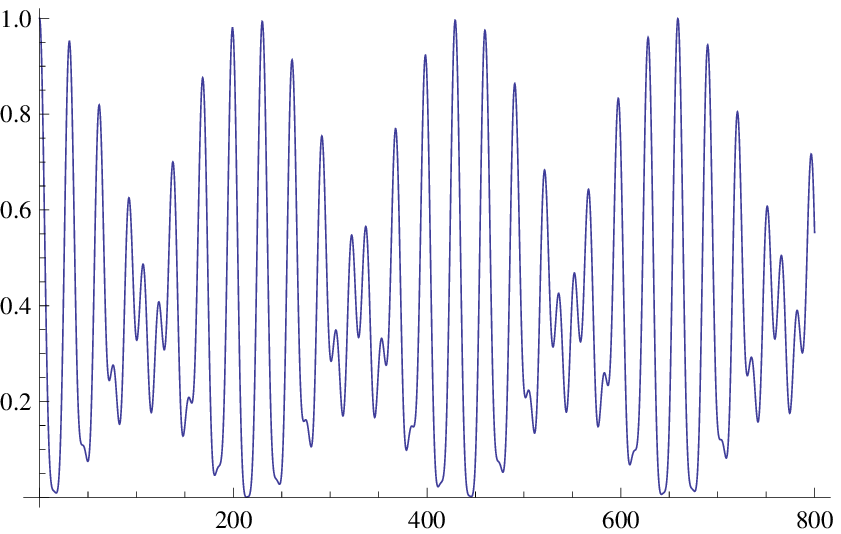}
\includegraphics[width=0.38\textwidth] {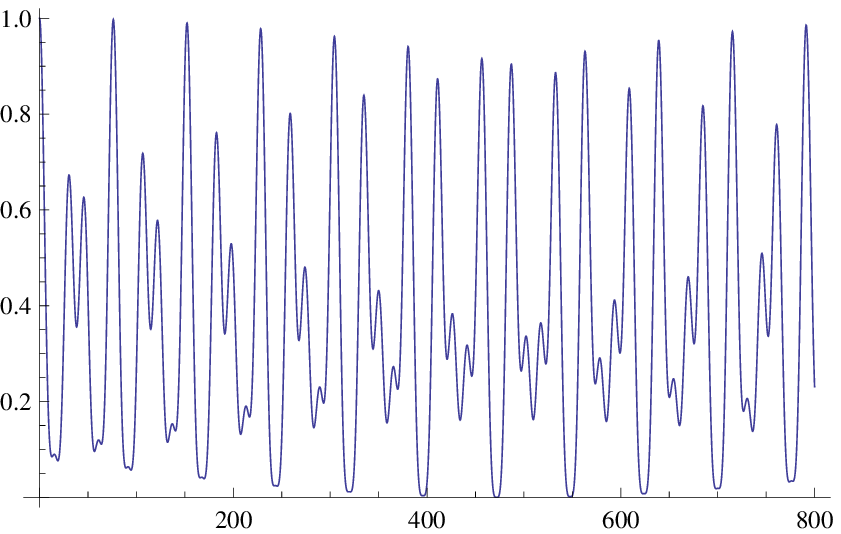}
 \includegraphics[width=0.38 \textwidth]{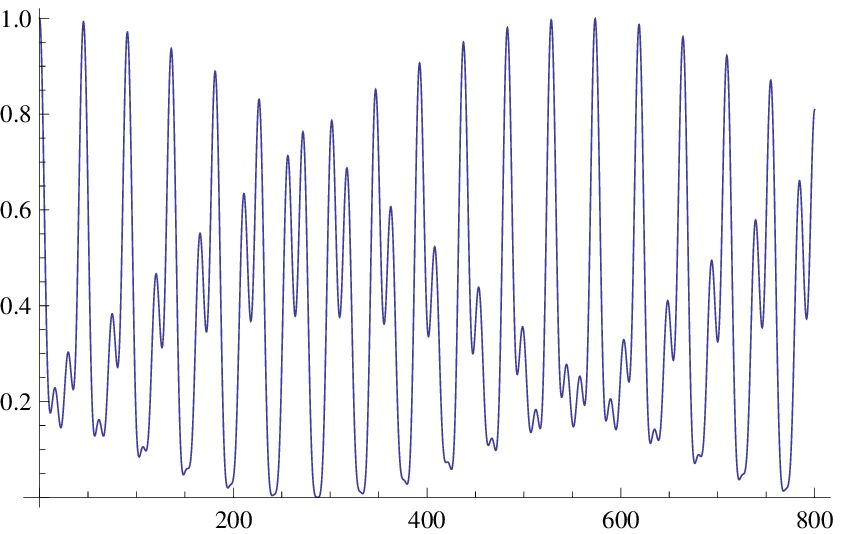}
 \caption{Schematic diagram of $P^1_N(t)$ with the four
parameters as $N=2,\ \frac{\omega_0}{\omega}=0.25,\ \beta=0.2,
a=0.2,\ 0,\ -0.2$ from the top to bottom respectively. The apparent
difference in these three figures strongly implies that the
parameter $a$ influences the qubits dynamically.}\label{fig11-3}
\end{figure}
In Ref.\cite{agar}, the authors discussed the special case where
$\Omega_{1N}\gg 2\beta^2$ with $a=0$. The other extreme case is that
$\Omega_{1N}\approx 0$: $a=0$ means that the spectrum of $H^0$ is
the same as that of $H$ in AA method, but they are different
whenever $a\ne 0$.

All eigen-values $\tilde{E}_N^m, \ m=0,\pm $ are influenced by the
parameter $a$ through the quantity $\tilde{T}=-2\beta^2+\frac{a}2
\frac{\omega_0}{\omega}(1-\exp{(-4\beta^2)}L_N(8\beta^2))$, so are
the eigenvectors $|\tilde{E}_N^{\pm}\rangle $. Of course, the
dynamics of the qubits will definitely  be differently from that of
the equal-level one.

\section{The Physical impact of the results}

The dynamics of the qubits is important for the real system. The
evolutionary behavior of the qubits depends  crucially on the
initial states, the initial states both for the qubits and the
quantum oscillator. Here we employ the initial state for the qubits
are $|1,m\rangle,\ m=0,\pm1$, which are applicable in the strong
coupling (see reference\cite{agar} for detail.). $|1,m\rangle,\
m=0,\pm1$ are different from the states $|m\rangle,\ m=1,\ 2,\ 3$.
Similarly, the natural initial states for the quantum oscillator are
the displaced number states or the displaced coherent states. In the
report, we treat the simplest case of the initial states
$|\Phi^N_m(0)\rangle=|1,m\rangle |N_m\rangle, \ m=0,\pm1$. We mainly
focus on the two kinds of  probabilities, that is, $P^m_N(t)$ for
the system remains unchanged and $T^N_{m\rightarrow l}(t)$  for it
transits to new states $|1,l\rangle |N_l\rangle,\ l\ne m$. It is
easy to obtain the following
\begin{eqnarray}
 P^1_N(t)&=&\frac14+\frac1{\tilde{L}_{N,+}^4}+\frac1{\tilde{L}_{N,-}^4} +\frac1{\tilde{L}_{N,+}^2}\cos \omega_{N,1}t
 \nonumber\\ &&+\frac1{\tilde{L}_{N,-}^2}\cos \omega_{N,2}t+\frac2{\tilde{L}_{N,+}^2\tilde{L}_{N,-}^2}\cos \omega_{N,0}t
\label{p1}, \nonumber \\
\end{eqnarray}
\begin{eqnarray}
&& P^0_N(t)=
\frac{\tilde{Y}_{N,+}^4}{\tilde{L}_{N,+}^4}+\frac{\tilde{Y}_{N,-}^4}{\tilde{L}_{N,-}^4}
+\frac{2\tilde{Y}_{N,+}^2\tilde{Y}_{N,-}^2}{\tilde{L}_{N,+}^2\tilde{L}_{N,-}^2}\cos
\omega_{N,0}t,  \nonumber \\
\end{eqnarray}
\begin{eqnarray}\tilde{T}_{1\rightarrow -1}^N(t)&=& \frac14+\frac1{\tilde{L}_{N,+}^4}+\frac1{\tilde{L}_{N,-}^4}
-\frac1{\tilde{L}_{N,+}^2}\cos \omega_{N,1}t \nonumber\\
&& -\frac1{\tilde{L}_{N,-}^2}\cos
\omega_{N,2}t+\frac2{\tilde{L}_{N,+}^2\tilde{L}_{N,-}^2}\cos
\omega_{N,0}t,  \nonumber \\
\end{eqnarray}
and
\begin{eqnarray}
&&\tilde{T}_{1\rightarrow
0}^N(t)=\frac{\tilde{Y}_{N,+}^2}{\tilde{L}_{N,+}^4}+\frac{\tilde{Y}_{N,-}^2}{\tilde{L}_{N,-}^4}
+\frac{2\tilde{Y}_{N,+}\tilde{Y}_{N,-}}{\tilde{L}_{N,+}^2\tilde{L}_{N,-}^2}\cos
\omega_{N,0}t, \nonumber \\
\end{eqnarray}
where
\begin{eqnarray} &&  \omega_{N,1}=\omega\left(\frac{4\Omega_{2N}-\tilde{T}_0+ \sqrt{\tilde{T}_0^2+8\Omega^2_{1N}}}2\right),
 \\ &&  \omega_{N,2}=\omega\left(\frac{4\Omega_{2N}-\tilde{T}_0- \sqrt{\tilde{T}_0^2+8\Omega^2_{1N}}}2\right),
   \\ &&\omega_{N,0}=\omega\sqrt{\tilde{T}_0^2+8\Omega^2_{1N}}\label{omega00}.
\end{eqnarray}

We see that the probabilities $P^0_N(t)$ oscillate with only one
frequency $\omega_{N,0}$, but the probabilities
$P^1_N(t)=P^{-1}_N(t)$ oscillate with three frequencies
 $\omega_{N,1}, \ \omega_{N,2},\ \omega_{N,0}$.  The non-equal level parameter $a$ changes the three
frequencies as well as the amplitude. For $a=0$ , the detailed
dynamic of the qubits is given in Ref.\cite{agar}. The special case
with $\beta^2\ll 8\Omega_{1N}^2,\ a=0$ is studied in
Ref.\cite{agar}, where $\omega_{N,1}=-\omega_{N,2}=\frac12
\omega_{N,0}=\sqrt2|\Omega_{1N}|$.

Figs.(\ref{fig11-3}) show the general trait for the qubit remaining
in its initial states $|1,\pm\rangle$ for different parameter
$a=0.2,\ 0, \ -0.2$. Obviously,  $P^1_N(t)$ is influenced
  by four parameters $\beta, a, N,
\frac{\omega_0}{\omega}$. In Ref.\cite{iris}, it is shown that the
coupling strength $\beta$ ranges from $0.01$ to $1$ for the
application of adiabatic approximation (weak coupling will not be
discussed here). From
Eqs.(\ref{e0})-(\ref{t0}),(\ref{p1}-\ref{omega00}), we see that the
parameter $a$ will come to action apparently whenever $\beta\approx
0.01-0.6$. As stated before, the qubits is equivalent to a two
qubits system and the non-equal-energy-level parameter $a$
represents the coupling strength between the two qubits. This shows
the coupling of the two qubits changes their dynamics considerately
in the range of $\beta\approx 0.1-0.6$ for the adiabatic
approximation method to be applied, and this is our limit on the
coupling parameter $\beta$

\section{Dynamics of the system}

Here we consider the dynamics of the composite system with different
initial conditions. Furthermore, we could also calculate the
probability for the two qubits stay in their other fully entangled
states. There are four fully entangled Bell states
\begin{eqnarray}
\Psi_{\pm}&=&\frac1{\sqrt2}(|\uparrow\downarrow\rangle\pm |\downarrow\uparrow\rangle)\\
\Phi_{\pm}&=&\frac1{\sqrt2}(|\uparrow\uparrow\rangle\pm
|\downarrow\downarrow\rangle).
\end{eqnarray}
It is easy to see that $|1,0\rangle=\Phi_-$, and the others are
related with the vectors $|1,1\rangle,\ |1,-1\rangle$. Due to the
facts that \[|2\rangle=\frac1{\sqrt2}(|1,1\rangle-|1,-1\rangle)\]
and \[|2\rangle=\frac1{\sqrt2}(|\uparrow\uparrow\rangle+
|\downarrow\downarrow\rangle)\] for the symmetrical case
$a=\kappa>0$ and
\[|2\rangle=\frac1{\sqrt2}(|\uparrow\uparrow\rangle-
|\downarrow\downarrow\rangle)\] for the symmetrical case
$a=-\kappa<0$,  We could write them as
\begin{eqnarray}
\Phi_-&=&|1,0\rangle\\
\Phi_{+}&=&\frac1{\sqrt2}(|1,1\rangle+|1,-1\rangle)\\
\Psi_{\pm}&=&|2\rangle=\frac1{\sqrt2}(|1,1\rangle-|1,-1\rangle).
\end{eqnarray}
Note that $\Psi_+,\ \Psi_-$ are vectors correspond to the positive
and negative sign of the parameter $a$ respectively. So the initial
fully entangled Bell states of qubits can be united to be written as
$|I_{\delta}\rangle=\frac1{\sqrt2}(|1,1\rangle+\delta |1,-1\rangle)$
($\delta=\pm1$) for $\Phi_+,\ \Psi_{\pm}$ and
$|I_0\rangle=|1,0\rangle$ for $\Phi_-$. We will consider the
dynamics of the system with the two qubits in $|I_{\pm 1}\rangle $
or $|I_0\rangle $ and quantum mode field in $ |\alpha\rangle$.

 \subsection{The qubits are initially  in states $|I_{\pm 1}\rangle $}

In first case,  the initial state for the two qubits is
$|I_{\delta}\rangle $ with $\delta=\pm1$ and quantum mode field in $
|\alpha\rangle$. Suppose $\beta<0.7$. Then in the adiabatic
approximation,  the qubits  will evolute into the states
$|I_{\bar{\delta}}\rangle $ with $\bar{\delta}=\pm1$ with
probability
\begin{widetext}
\begin{eqnarray}
 &&P(\delta,\bar{\delta},\alpha,t)=\frac12+\sum_{N=0}^{\infty}\frac{\bar{\delta}\delta}{N!}(\alpha^2-\beta^2)^Ne^{-(\alpha^2+\beta^2)}\langle
 N_{-1}|N_1\rangle\nonumber\\ &&-\frac12\sum_{N=0}^{\infty}\bigg(p(N,\alpha+\beta)+p(N,\alpha-\beta)
 +\frac{2\delta}{N!}(\alpha^2-\beta^2)^Ne^{-(\alpha^2+\beta^2)}\bigg)(1+\langle
 N_{-1}|N_1\rangle)
 \frac{\Omega^2_{1N}}{\tilde{T}_0^2+8\Omega^2_{1N}}\bigg(1-\cos
(\omega_{N,0}t)\bigg), \nonumber \\
\label{p1pm to p1pm}
\end{eqnarray}
\end{widetext}
where \begin{eqnarray}
p(N,\alpha)=\frac{e^{-\alpha*\alpha^*}|\alpha|^{2N}}{N!}
\end{eqnarray} is the probability of quantum field state
$|\alpha\rangle$ in its number state $|N\rangle$ and $\delta=\pm1,\
\bar{delta}=\pm1$. From the above Eq.(\ref{p1pm to p1pm}), we see
that the initial entangled states $|I_{\pm1}\rangle$  of the two
qubits will have large possibility ($p(\delta,-\delta,\alpha,t)$
being around $\frac12$ ) to evolute into the states
$|I_{\mp1}\rangle$. So it is hard for the two qubits to remain its
entangled states. In the following subsection, we will check that
for the initial state $|I_0\rangle$.

\subsection{The qubits are initially  in states $|I_{0}\rangle $}

suppose the qubits is in the state $|I_0\rangle$ initially, while
the oscillator naturally stays in its coherent state
$|\alpha\rangle$. This is a very general state for a quantum
processing system.
 The system will evolute accordingly and the probability for the qubits remain in its initial state is
 \[P_0(\alpha)=1-T(\alpha,t),\]
where $T(\alpha,t)$ is the probability of the two qubits transiting
to other non-entangled states and is  given as
\begin{eqnarray}
T(\alpha,t)&=&\sum_{N=0}^{\infty}2p(N,\alpha)\frac{\tilde{Y}_{N,+}^2}{\tilde{L}_{N,+}^4}\bigg(1-\cos
(\omega_{N,0}t)\bigg),\label{t0-1}\end{eqnarray} where $p(N)$ is the
probability of $N$ photons in the coherent state $|\alpha\rangle$.
In the large quantity $|\alpha|^2\gg1$ approximation, the quantity
$P(t)$ could be simplified greatly due to the fact \begin{eqnarray}
p(N,\alpha)&=&\frac{e^\frac{{-(N-|\alpha|^2)^2}}{2|\alpha|^2}}{\sqrt{2\pi|\alpha|^2}}.
\end{eqnarray}

\begin{figure}[ht]
\centering
\includegraphics[width=0.38\textwidth]{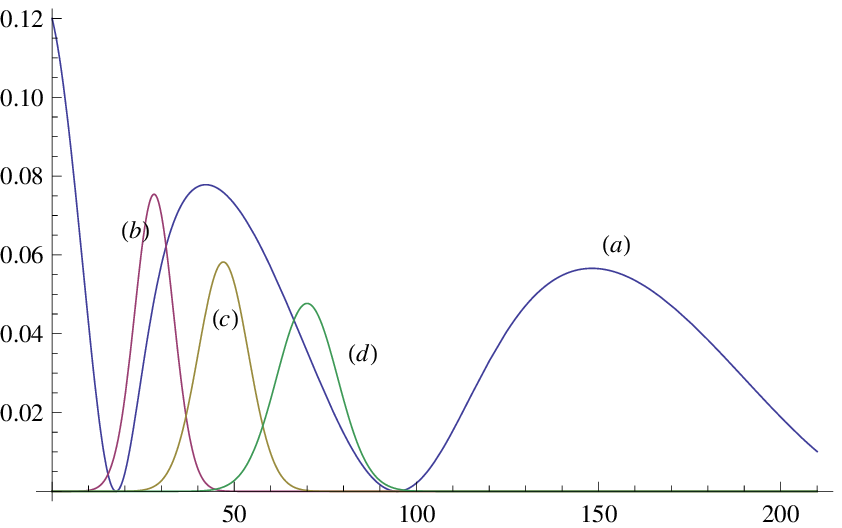}
\includegraphics[width=0.38\textwidth]{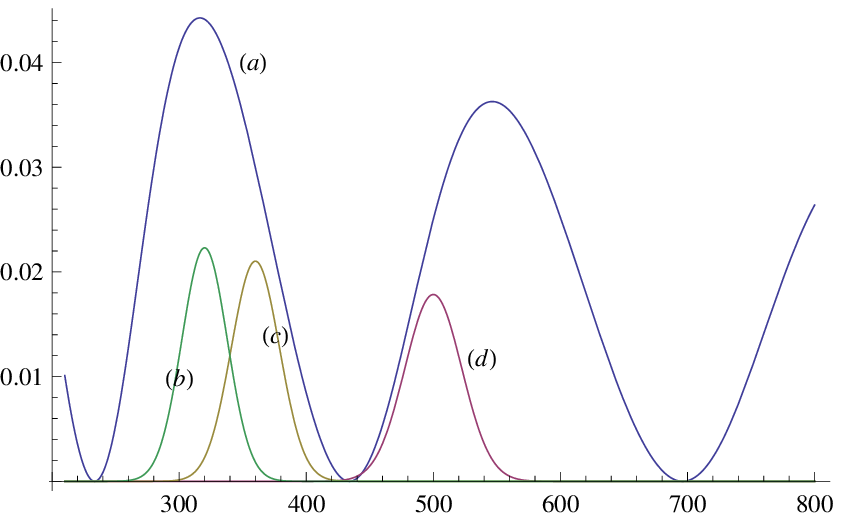}
\caption{The function $B(N,t)$, $P(N,\alpha)$ with
$|\alpha|^2=320,360,500$. lines (a),(b),(c),(d) correspond to
$B(N,t)$ with parameters $a=-0.8,\ \beta=0.2,\
\frac{\omega_0}{\omega}=0.24$ and $P(N,\alpha)$ with
$|\alpha|^2=28,47,70$ in the first figure and (b),(c),(d) lines
correspond to $B(N,t)$ with parameters $a=-0.8,\ \beta=0.2,\
\frac{\omega_0}{\omega}=0.24$ and $P(N,\alpha)$ with
$|\alpha|^2=320,360,500$ in the second figure .}\label{fig7}
\end{figure}

We denote the term $2\frac{\tilde{Y}_{N,+}^2}{\tilde{L}_{N,+}^4}$ as
$B(N)$ in the above equation and  will show the general properties
of $B(N)$ by some special parameters in Fig.(\ref{fig7}).
Fig(\ref{fig7}) also gives some examples concerning the
$p(N,\alpha)$ in case of $|\alpha|^2\gg1$. Under these assumption of
rapid falling to zero of $p(N,\alpha)$ as $N$ deviated from its
average $|\alpha|^2$, we could safely approximate $B(N)$ in
Eq.(\ref{t0-1}) as
\[B(N)\approx b_0+b_1(N-\bar{n})+b_2(N-\bar{n})^2\] where $\bar{n}=[|\alpha|^2]$
is the integer part of $|\alpha|^2$
 and \[b_0=2\frac{\tilde{Y}_{\bar{n},+}^2}{\tilde{L}_{\bar{n},+}^4},\ b_1=\bigg[\frac {dB(N)}{dN}\bigg]_{N=\bar{n}},\
 b_2=\bigg[\frac {d^2B(N)}{2dN^2}\bigg]_{N=\bar{n}}.\]

It is easy to calculate $T(\alpha,t)$ as two parts
\begin{eqnarray} T(\alpha,t)&=&T_1(\alpha)-T_2(\alpha,t)\\
T_1(\alpha)&=&\sum_{N=0}^{\infty}B(N)
\frac{e^\frac{{-(N-|\alpha|^2)^2}}{2|\alpha|^2}}{\sqrt{2\pi|\alpha|^2}}=b_0+b_2\bar{n}\\
T_2(\alpha,t)&=& \sum_{N=0}^{\infty}B(N)
\frac{e^\frac{{-(N-|\alpha|^2)^2}}{2|\alpha|^2}}{\sqrt{2\pi|\alpha|^2}}\cos
(\omega_{N,0}t).
\end{eqnarray} 
In the assumption that $|\alpha^2|\gg
1$  and Gauss form of $p(N,\alpha)$, it is reasonable to extend the
summation in $T_2(\alpha,t)$ from $0$ to $-\infty$. Then the use of
Poisson summation formula gives
\begin{eqnarray} 
T_2(\alpha,t)&=&\sum_{k=-\infty}^{+\infty}\bar{g}(k,t)
\end{eqnarray}
\begin{eqnarray}
\bar{g}(k,t)&=& \int_{-\infty}^{+\infty}B(N)
\frac{e^\frac{{-(N-|\alpha|^2)^2}}{2|\alpha|^2}}{\sqrt{2\pi|\alpha|^2}}\cos
(\omega_{N,0}t)e^{i2\pi kN} dN.  \nonumber \\
\end{eqnarray} 

Generally,
$\omega_{N,0}t$ could be simplified as
\[\omega_{N,0}t=\omega_{\bar{n},0}t+c_1(N-\bar{n})+c_2(N-\bar{n})^2\]
with \[c_1=\bigg[\frac {d\omega_{N,0}t}{dN}\bigg]_{N=\bar{n}},\
c_2=\bigg[\frac {d^2\omega_{N,0}t}{2dN^2}\bigg]_{N=\bar{n}}.\] 
So we have
\begin{eqnarray}
\bar{g}(k,t)(t)&=&A(k,t)\cos\theta_1(k,t)\label{gbar1},\\
        A(k,t)&=& \frac{e^{-\frac{\bar{n}(2\pi
k+c_1t)^2}{(1+4\bar{n}^2c_2^2t^2)^{\frac12}}\cos\theta(t)}}{\bigg(1+4\bar{n}^2c_2^2t^2\bigg)^{\frac14}}\bigg(b_0
+\gamma\bigg),\\
\theta_1(k,t)&=&\theta_1(t)=(\omega_{\bar{n},0}t+\frac{\theta(t)}2+2\pi
k\bar{n})\nonumber\\ 
&& -\frac{\bar{n}(2\pi
k+c_1t)^2}{(1+4\bar{n}^2c_2^2t^2)^{\frac12}}\sin\theta(t),\\
\tan\theta(t)&=&2\bar{n}c_2t.\label{gbar2} \\
\gamma&=&b_2(\frac{\bar{n}}{(1+4\bar{n}^2c_2^2t^2)^{\frac12}}-\frac{(2\pi k+c_1t)^2\bar{n}^2}{(1+4\bar{n}^2c_2^2t^2)}). \nonumber \\
\label{gbar3}
\end{eqnarray}

It is clear that $T_2(\alpha,t)$ exhibits the collapse and revival
phenomena with its $k-th$ term being $\bar{g}(k,t)$. It is more
useful to delineate them in some special cases: one case that when
$c_1\ne 0$ with $c_2= 0$, and the other extreme case that $c_1=0$
and $c_2\ne 0$. In the first case, we obtain that
\begin{equation}\label{111}
  A(k,t)=  e^{-\bar{n}(2\pi
k+c_1t)^2}\bigg(b_0 +b_2(\bar{n}-(2\pi k+c_1t)^2\bar{n}^2)
       \bigg)
\end{equation}
The revival time \[t_{rev}(k)=2\pi|\frac{k}{c_1}|\] and the height
for its amplitude is \[A(k,t_{rev})=b_0 +b_2\bar{n}=T_1,\] which is
constant in contrast to the decreasing height as time goes in
Ref.\cite{agar}.

In the second case, it is easy to see that \begin{eqnarray}
        A(k,t)&=& \frac{e^{-\frac{4\bar{n}\pi^2
k^2}{(1+4\bar{n}^2c_2^2t^2)}}}{\bigg(1+4\bar{n}^2c_2^2t^2\bigg)^{\frac14}}\nonumber\\
& *& \bigg(b_0
+\frac{\bar{n}b_2}{(1+4\bar{n}^2c_2^2t^2)^{\frac12}}-\frac{4\pi^2
k^2\bar{n}^2b_2}{(1+4\bar{n}^2c_2^2t^2)}
       \bigg), \nonumber \\
       \end{eqnarray}
where the fact 
\[\tan\theta(t)=2\bar{n}c_2t,\ \
\cos\theta(t)=\frac1{(1+4\bar{n}^2c_2^2t^2)^{\frac12}}\] are used.
Obviously, there is no revival phenomena in $A(k,t)$ in this case.
$A(k,t)$ also will generally decreases as time goes except for the
initially  irregular transiting change.

This two cases are not possible absolutely. Anyway, $c_2$ may be
very small with $c_1\gg c_2$, and this case is close to the first
case, where collapse and revival appear in $T_2(\alpha,t)$. However,
the small and non-equal-zero quantity  $c_2$ contributes both the
decreasing heights of the revival amplitude, as is shown in
Eq.(\ref{gbar1})-(\ref{gbar3}) and the broadening  of revivals with
the time growing. The broadening of revivals  also makes the
collapse interval shorter and shorter until its disappearing.

\begin{figure}[ht]
\centering
\includegraphics[width=0.38\textwidth]{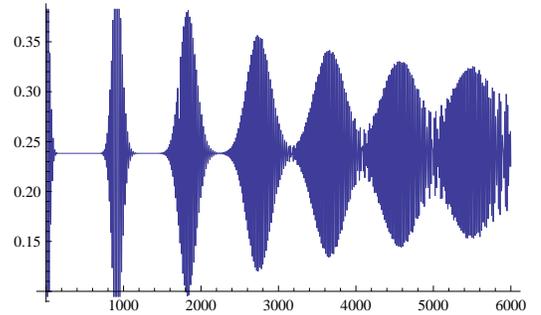}
\caption{The transition  $2T(\alpha,t)$ with $|\alpha|^2=14,\
a=-0.48,\ \beta=0.102,\ \frac{\omega_0}{\omega}=0.21$.}\label{fig80}
\end{figure}

Similarly, $c_2\gg c_1$ and $c_1\ne 0$ means the revival time gap is
greater than that in the first case.  All these features all shown
in Fig.(\ref{fig80}).

The above discussion might be limited or not applicable in the
following case, where $p(N,\alpha)$ might be decrease not rapid
enough that the approximation $B(N)\approx
b_0+b_1(N-\bar{n})+b_2(N-\bar{n})^2$ and $
\omega_{N,0}t=\omega_{\bar{n},0}t+c_1(N-\bar{n})+c_2(N-\bar{n})^2 $
can fails to hold. Then new detailed approximation must be added. As
this is seldom, we just stop here.

\section{The preservation of entanglement of two qubits}
From the previous section, we see that the dynamics of the Rabi
model is much more complicated than that of its RWA counterpart  JC
model.
\begin{figure}[ht]
\centering
\includegraphics[width=0.38\textwidth]{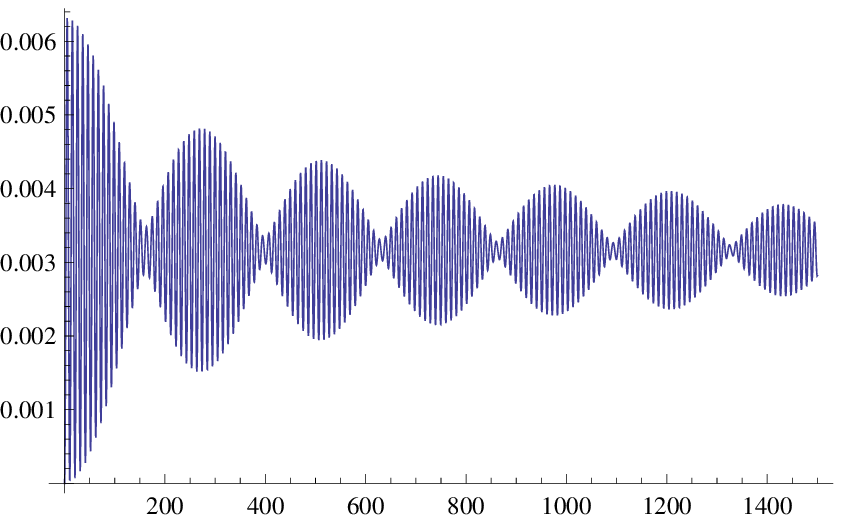}
\includegraphics[width=0.38\textwidth]{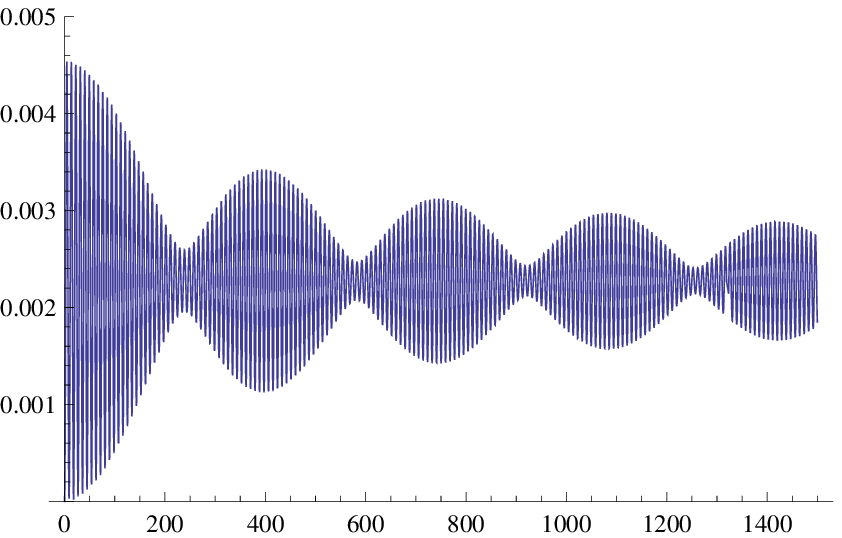}
\caption{The function $T(\alpha,t)$ becomes very small with
$|\alpha|^2=36,\ a=-0.8,\ \beta=0.5067,\
\frac{\omega_0}{\omega}=0.22$ for upper figure and $|\alpha|^2=55,\
a=-0.6,\ \beta=0.5599,\ \frac{\omega_0}{\omega}=0.24$  for lower
figure. }\label{fig8}
\end{figure}
To initial entangled state $|I_0\rangle$ of the two qubits with
control field mode in its coherent state, its evolution involves on
the various parameters in Eq.(\ref{t0-1}). It depends on the number
$N$ in an extremely nonlinear and intricate way. The kind perplexity
makes it hard to study the Rabi model, never the less, it also
provide the opportunity to preserve the entanglement of the two
qubits by careful choice of the appropriate parameters. Because the
coherent state has the probability of Poisson distribution, which
will be approximated by a Gauss distribution if the average number
$|\alpha|^2$ is large enough. The intricacy of the Rabi model could
be utilized to make the quantity
$\frac{\tilde{Y}_{N,+}^2}{\tilde{L}_{N,+}^4}=B(N)$ in
Eq.(\ref{t0-1}) extremely small when $N$ is in the neighborhood of
$|\alpha|^2$ by some selection of the appropriate parameters, which
will guaranty the initial state of the qubits unchanging. This is
shown in Fig.(\ref{fig8}).  It can be easy to see that whenever we
select the parameters appropriate, for example, the parameters as
$|\alpha|^2=55,\ a=-0.6,\ \beta=0.5599,\
\frac{\omega_0}{\omega}=0.24$, the Bell state
$|I_0\rangle=|1,0\rangle =\frac1{\sqrt2}(|\uparrow\uparrow\rangle
-|\downarrow\downarrow\rangle)$ have the probability about
$1-0.005=\frac{99.5}{100}$ to remain unchanged\cite{tian}.

$\frac{\tilde{Y}_{N,+}^2}{\tilde{L}_{N,+}^4}$ in Eq.(\ref{t0-1})
being small in the neighborhood of $N=|\alpha^2|$ is crucial for
$T(\alpha,t)\approx 0$. So we will select the zeros ($N_1,N_2,\cdots
$) of $\Omega_{1N}$ as possible parameters for $|\alpha|^2$ under
definite quantities $\beta,\ a$. Then $|\tilde{T}_0|^2=|
-2\beta^2+\frac{a}2 \frac{\omega_0}{\omega}+\Omega_{2N}|^2$ larger
around zeros of $\Omega_{1N}$ will advantage $T(\alpha,t)\approx 0$.
As a result, $a$ negative and $\Omega_{2N}$ negative around zeros of
$N=|\alpha|^2$ of $\Omega_{1N}$ are keys to make $T(\alpha,t)\approx
0$, that is, to keep the entangled state $|I_0\rangle$ unchanged.

 There is an alternative method for the realization of $T(\alpha,t)\approx
 0$. One could first determine the average number $T(\alpha,t)\approx 0$ of the coherent state of  the control
 field, then chooses $\beta$ and $a$ by similar requirement that $|\tilde{T}_0|^2=| -2\beta^2+\frac{a}2
\frac{\omega_0}{\omega}+\Omega_{2N}|^2$ as larger as possible around
$N=|\alpha|^2$.

The parameter $a$ is connected with the inter-qubit coupling
strength $\kappa$ as $a=\pm \kappa$  in symmetric and asymmetric
transition cases respectively. Study also shows that the parameter
$a$ negative is favorable for $|T(\alpha,t)$ approaching zero, as
Fig.(\ref{fig8}) exhibits. So the inter-qubit coupling is in favor
of preservation of the initial entanglement, especial with
asymmetrical transition case ($a<0$).

All the others' Bell states have not this nice property because
there is a simple factor $\frac12$ in quantities
$P(\delta,-\delta,\alpha,t)$ in Eq.(\ref{p1pm to p1pm}). Never the
less, the preservation of the entangled Bell state $|I_0\rangle$ is
still useful for its application in quantum information process.
Also, the complex formula of $T(\alpha,t)$ make the appropriate
choice of the parameters much easier and will beneficial to the
experiment application.

 In summary, coupled strongly with a quantum mode field, the two
qubits' dynamics is influenced by three parameters $ \beta,\
\frac{\omega_0}{\omega},\  a$ and the initial conditions in a very
complicated form. We investigate the evolution of the four Bell
entangled states with the control mode in its coherent state. Three
out of the four Bell states will become the combination of the four
Bell states and can not remain in their initial entangled states.
Nevertheless, the above mentioned complexity unexpectedly promotes
our ability to preserve the entanglement of the two qubits in  one
Bell state $|I_0\rangle=|1,0\rangle
=\frac1{\sqrt2}(|\uparrow\uparrow\rangle
-|\downarrow\downarrow\rangle)$, that is, it could remain in its
initial states by suitable choice of the controlled parameters. It
is shown that the
 parameter $a$ negative is more favorable for the maintaining the
state $|I_0\rangle=|1,0\rangle$.
 These results will be useful for the information process.

\acknowledgments  The work was partly supported by the Major State
Basic Research Development Program of China (973 Program:
No.2010CB923202) and the National Natural Science of China (No.
10875018).

\end{document}